\documentclass[aps,pre,twocolumn,groupedaddress,showpacs,showkeys]{revtex4}

\usepackage{graphicx}
\usepackage{dcolumn}
\usepackage{bm}
\usepackage{amssymb}

\graphicspath{{Fig/}}

\usepackage{color}

\newcommand{\alkor}[1]{\textcolor{red}{#1}}
\renewcommand{\alkor}[1]{#1}

\begin{document}

\title{Computation of the spectrum of spatial Lyapunov exponents for the spatially
extended beam-plasma systems and electron-wave devices}

\author{Alexander E.~Hramov}
\author{Alexey~A.~Koronovskii}
\author{Vladimir~A.~Maximenko}
\author{Olga~I.~Moskalenko}
%
\affiliation{Faculty of Nonlinear Processes, Saratov State
University, Astrakhanskaya str., 83, Saratov, 410012, Russia}
\date{\today}

\begin{abstract}
\noindent  The spectrum of Lyapunov exponents is powerful tool for the
analysis of the complex system dynamics. In the general
framework of nonlinear dynamical systems a number of the numerical technics have been developed to obtain
the spectrum of Lyapunov exponents for the complex temporal behavior of the systems with a few degree of freedom. Unfortunately,
these methods can not apply directly
to analysis of complex spatio-temporal dynamics in plasma devices which are characterized by the infinite phase space,
since they are the spatially extended active media. In the present paper we propose the method for the calculation of the spectrum of the spatial Lyapunov
exponents (SLEs) for the spatially extended beam-plasma systems.
The calculation technique is applied to the analysis of chaotic spatio-temporal oscillations in three different
beam-plasma model: (1) simple
plasma  Pierce diode, (2) coupled Pierce diodes, and (3)
electron-wave system with backward electromagnetic wave.
We find an excellent agreement between the system dynamics
and the behavior of the spectrum of the spatial Lyapunov exponents. Along with
the proposed method, the possible problems of SLEs calculation are also
discussed. It is shown that for the wide class of the spatially extended
systems the set of quantities included in the system state for SLEs calculation can be
reduced using the appropriate feature of the plasma systems.
\end{abstract}

\pacs{05.45.-a, 05.10.-a}
\keywords{spectrum of Lyapunov exponents, spatially extended
systems, beam plasma systems, electron wave media, Pierce diode,
transverse field backward wave oscillator, reference system
state.}

\maketitle

\section*{Introduction}
\label{Sct:Intro}

Nonlinear dynamics and chaos in plasma devices and microwave systems
with interacted electron and electromagnetic waves  is often
observed in bounded plasmas and electron and ions streams, where the
nonlinearity and plasma instabilities are combined with dissipation
caused by current flow to the boundary electrodes. Therefore, the
development of the methods of analysis of spatially extended
beam-plasma systems is a very important problem due to such
applications of chaos theory in plasma physics as the identification
of chaotiс spatial-temporal oscillations in microwave and plasma
devices (e.g. free electron lasers, pasotrons, vircators, etc)
\cite{Bogomolov:1985_FEL, PhysRevA.43.5541, Goryashko:2009_FEL},
chaos control and suppression of turbulence
\cite{Klinger:2001_PlasmaChaos, hramov:013123, goud:032307,
toufen:012307, klinger:1997}, and chaotic synchronization in the
plasma and microwave electronic systems
\cite{Ticos:2000_PlasmaDischarge, PhysRevE.63.056401,
Hramov:2005_Chaos_BWO, Filatov:2006_PierceDiode_PLA}.

The Lyapunov exponents are powerful and useful tool for the
complex system dynamics to be analyzed. They are used widely to
characterize the systems with small number of degrees of freedom in the different
fields of science~\cite{Thamilmaran:2006_SNAinCircuit, Porcher:2001_LEinMedicine, Karakasidis:2007_MolecularDynamicsPRE, Dunki:2000_LEEstimation}, such as
physics, physiology and medicine, molecular
dynamics, astronomy, etc.

At the same, the majority of the plasma systems are characterized by the infinite phase space, since they are the spatially extended active media. Unfortunately, the direct application of the Lyapunov exponent calculation technique developed for the system with the small number of degrees of freedom to the objects of the beam-plasma and electron-wave systems being the spatially extended systems is problematic. The main causes of the this problem are the following: (i) the phase space of the spatially distributed system is infinite; (ii) the number of the Lyapunov exponents is also infinite; (iii) the \emph{system state} (defined for each type of the system under study in its own way) must be used instead of the finite-dimensional vector; (iv) the orthogonalizaiton and normalization procedures based on Gram--Schmidt orthonormalization  must be modified to be applicable for the \emph{states}.

Nevertheless, the scientists try to use the Lyapunov exponents for
the spatially extended systems due to the great efficiency of this
tool. Mainly, all attempts are reduced to the use of the
modifications of the standard methods developed for the finite
dimensional systems (e.g., the estimation of the highest Lyapunov
exponent from time series or the representation of the spatially
extended system as the finite dimensional one with the high
dimension by means of the discretization). Unfortunately, these
approaches do not take into account the distinguishing
characteristics of the spatially extended systems and have serious
limitations connected with the high dimension of the phase space of
the discretized system

Recently, the successful attempt to compute the highest Lyapunov
exponent taking into account the core features of the spatially
extended systems has been made for plasma system
models~\cite{Filatov:2006_PierceDiode_PLA, hramov:013123}. In the
present work we develop the technique of the spectrum of the spatial
Lyapunov exponents (SLEs) calculation for the distributed
beam-plasma systems and electron-wave devices. The structure of the
article  is the following. In Sec.~\ref{sct:SLEsTheory} the general
idea of the proposed approach is given. The next
Sec.~\ref{sct:LEsForPierceDiode} contains the results concerning the
computation of the spectrum of spatial Lyapunov exponents for the
spatially extended beam plasma system. As the model system under
study the Pierce diode has been selected, with the spectrum of
Lyapunov exponents being calculated both for autonomous oscillator
and for two coupled plasma diodes. To illustrate the generality and
efficiency of the proposed method, in Sec.~\ref{Sct:Backward wave
oscillator} the SLEs have been also calculated for the transverse
field backward wave oscillator being the classical models of the
electrons-wave systems. The problem concerning the selection of the
system state for the SLEs computation is discussed in
Sec.~\ref{sct:SystemStateSelection}. The final remarks are given in
Conclusion.

\section{Theory of the spatial Lyapunov exponents}
\label{sct:SLEsTheory}

The main idea of the computation of the spectrum of the spatial
Lyapunov exponents is the following. Let
\begin{equation}\label{eq:Evolution}
\hat{L}(\mathbf{U}(x,t))=0
\end{equation}
be an operator determining the evolution of the spatially extended
system under study, $\mathbf{U}(x,t)$ is the system state which (in
the general case) depends on the spatial coordinate $x\in[0,L]$ and
time $t$.

The state $\mathbf{U}(x,t)$ plays the role of the reference point to
study the evolution of the small perturbations. To compute $N$
largest spatial Lyapunov exponents we consider the set of
perturbations $\mathbf{V}_i(x,t)$, $i=1,\dots,N$ which are
orthogonal
\begin{equation}\label{eq:OrtCond}
(\mathbf{V}_i,\mathbf{V}_j)=\left\{
\begin{array}{ll}
1, & i=j\\
0, & i\neq j\\
\end{array}\right.
\end{equation}
where $(\mathbf{U},\mathbf{V})$ is the scalar product. If the system
state is the $N_d$-dimensional vector
$\mathbf{U}=(u_1,\dots,u_{N_d})^T$ or real function  $\mathbf{U}=u(x,t)$
defined in $\mathbb{L}^2$ the definition of the scalar product is
\begin{equation}
(\mathbf{U},\mathbf{V})=\sum\limits_{i=1}^{N_d}u_i(t)v_i(t)
\end{equation} or
\begin{equation}
(\mathbf{U},\mathbf{V})=\int\limits_{L}u(x,t)v(x,t)\,dx,
\end{equation}
respectively. At the same time, the state of the spatially extended
beam-plasma system should be usually defined in a more complicated way, and, as
a consequence, the definition of the scalar product is also more
complex. E.g., if the system state is the $N_d$-dimensional
vector-function ${\mathbf{U}(x,t)=(u_1(x,t),\dots,u_{N_d}(x,t))^T}$
the scalar product should be defined as
\begin{equation}\label{eq:ScalarProductVectorfunction}
(\mathbf{U},\mathbf{V})=\sum\limits_{i=1}^{N_d}\int\limits_{L}u_i(x,t)v_i(x,t)\,dx.
\end{equation}

Additionally to the orthogonality requirement~(\ref{eq:OrtCond}) all
perturbations $\mathbf{V}_i(x,t)$ must also obey the normalization
condition
\begin{equation}\label{eq:NormCond}
||\mathbf{V}_i(x,0)||=1,
\end{equation}
where $||\mathbf{V}||=\sqrt{(\mathbf{V},\mathbf{V})}$.

The set of the perturbations $\mathbf{V}_i(x,t)$ fulfilling
requirements~(\ref{eq:OrtCond}) and (\ref{eq:NormCond}) may be
obtained with the help of the Gramm-Schmidt orthonormalization method
\begin{equation}
\left\{
\begin{array}{l}
\displaystyle
\mathbf{V}_i=\frac{\mathbf{\tilde V}_i(x,0)}{||\mathbf{\tilde V}_i(x,0)||}=\frac{\mathbf{\tilde V}_i(x,0)}{\sqrt{(\mathbf{\tilde V}_i(x,0),\mathbf{\tilde V}_i(x,0))}}\\
\displaystyle\mathbf{\tilde V}_1(x,0)=\varphi_1(x)\\
\displaystyle\mathbf{\tilde V}_{i+1}(x,0)=\varphi_{i+1}(x)-\sum\limits_{k=1}^{i}(\mathbf{V}_k,\varphi_{i+1})\mathbf{V}_k(x,0),\\
\end{array}\right.
\end{equation}
${i=1,2,\dots,N-1}$, where $\varphi_1(x)$, $\varphi_2(x)$, \dots,
$\varphi_N(x)$ is the set of the linearly independent arbitrary
states defined for the system under study.

To compute the spectrum of the spatial Lyapunov exponents of the
spatially extended system the behavior both of the state
$\mathbf{U}(x,t)$ of the system under study and all perturbations
$\mathbf{V}_i(x,t)$ (${i=1,2,\dots,N-1}$) has to be considered. The
evolution of the system state  $\mathbf{U}(x,t)$ is defined by
Eq.~(\ref{eq:Evolution}), whereas the development of the
perturbations $\mathbf{V}_i(x,t)$ of the reference state is
described by the linearization
\begin{equation}\label{eq:PerturbationEvolution}
\partial\hat{L}(\mathbf{U}(x,t),\mathbf{V}_i(x,t))=0
\end{equation}
of the evolution operator~(\ref{eq:Evolution}) applied in the
vicinity of the state $\mathbf{U}(x,t)$.

After the time interval with the length $T$ the obtained set of the
perturbations $\mathbf{V}_i(x,T)$ must be orthogonalized and normalized
with the help of the Gramm--Shmidt method again, with the set of the
functions $\varphi_i(x)$ being the considered perturbations
$\mathbf{V}_i(x,T)$. In other words,
\begin{equation}\label{eq:PerturbEvolution}
\varphi_i(x)=\mathbf{V}_i(x,T).
\end{equation}

The described algorithm for spatially extended system must be repeated many times. After $M$
iterations the Lyapunov sums are calculated
\begin{equation}\label{eq:SLEsums}
S_i=\sum\limits_{j=1}^M \ln||\mathbf{\tilde V}_i(x,jT)||,
\end{equation}
where the perturbations before renormalization but after
orthogonalizaiton are used. The spatial Lyapunov exponents are
estimated as
\begin{equation}\label{eq:SLE}
\Lambda_i=\frac{S_i}{MT}.
\end{equation}

\section{Spectrum of spatial Lyapunov exponents for Pierce diode}
\label{sct:LEsForPierceDiode}

To illustrate the efficiency of the proposed method for the
computation of spatial Lyapunov exponents we consider the behavior
of autonomous and coupled Pierce diodes \cite{Pierce:1944} being the
classical models of beam-plasma systems, demonstrating the complex
spatio-temporal oscillations including the chaotic
ones~\cite{Pierce:1944, Godfrey:1987, Matsumoto:1996}.

\subsection{Pierce diode}
\label{Sct:PierceDiode}

\begin{figure}[tb]
\centerline{\scalebox{0.3}{\includegraphics{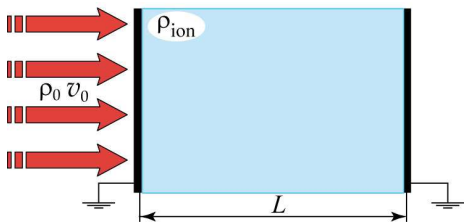}}}
\caption{(Color online) Schematic diagram of Pierce diode}
\label{fgr:Fig1}
\end{figure}

A Pierce diode~\cite{Pierce:1944, Godfrey:1987, Matsumoto:1996,
TrueAeh:2003_microwave_electronics_1engl} is schematically
illustrated in Fig.\,\ref{fgr:Fig1}, and consists of two infinite
parallel plains pierced by a mono-energetic electron beam. Grids are
grounded, with the distance between them being $L$. The charge
density $\rho_0$ and electron velocity $v_0$ are constant at the
system input. The region between two plains is uniformly filled
by neutralizing stationary ions, with the density
\alkor{$|\rho_\mathrm{ion}|$} being equal to the non-perturbed
electron beam density $|\rho_0|$.

The dimensionless Pierce parameter $ \alpha=\omega_pL/v_0$
determines the dynamics of the system (here $\omega_p$ is the
electron beam plasma frequency, $v_0$ is the non-perturbed electron
velocity, $L$ is the distance between the diode plains). With
$\alpha>\pi$ the so-called Pierce instability is developed in the
system and a virtual cathode is formed in the electron beam
\cite{Matsumoto:1996}. A detailed analysis of
the Pierce instability is provided in Ref.~\cite{Kuhn:1994_PierceInstability}.

Here we focus our attention on a narrow
range of  Pierce parameter near $\alpha\sim3\pi$ where the increase of the
instability is suppressed by the nonlinearity, and a reflectionless
regime takes place in the electron beam
\cite{Matsumoto:1996,TrueAeh:2003_microwave_electronics_1engl}. In
this case the system behavior may be described by the fluid
equations~\cite{TrueAeh:2003_microwave_electronics_1engl,
Godfrey:1987, Matsumoto:1996}, that give rise to various types of
beam-plasma chaotic oscillations \cite{Godfrey:1987, Kuhn:1990,
Matsumoto:1996, hramov:013123}.

The dynamics of the electron flow in the Pierce diode (in the fluid electronic
approximation) is described by the self-congruent system of
dimensionless Poisson, continuity and momentum equations
\begin{equation}\label{eq:diodPierce3}
\frac{\partial^2\varphi}{\partial x^2}=-\alpha^2(\rho-1),
\end{equation}
\begin{equation}\label{eq:diodPierce2}
\frac{\partial\rho}{\partial t}+ v\frac{\partial \rho}{\partial
x}+\rho\frac{\partial v}{\partial x}=0,
\end{equation}
\begin{equation}\label{eq:diodPierce1}
\frac{\partial v}{\partial t}+v\frac{\partial v}{\partial
x}=-\frac{\partial \varphi}{\partial x},
\end{equation}
with the boundary conditions
\begin{equation}\label{eq:diodPierce_usloviya}
v(0,t)=1,\quad \rho(0,t)=1,\quad\varphi(0,t)=\varphi(1,t)=0,
\end{equation}
where $\varphi(x,t)$ is the dimensionless potential of the electric
field, $\rho(x,t)$ and $v(x,t)$ are the dimensionless density and
velocity of the electron beam ($0\leq x \leq1$), respectively. The
dimensional variables ($\varphi'$, $\rho'$, $v'$, $x'$, $t'$) are
connected with the dimensionless ones as $\varphi '=
({v^2_0/\eta})\varphi$, $\rho ' =\rho_0\rho$, $v'=v_0 v$, $x'=L x$,
$t'=({L/v_0})t,$ where $\eta$ is the specific electron charge, $v_0$
and $\rho_0$ are the non-perturbed velocity and density of the
electron beam.

The system~(\ref{eq:diodPierce3})-(\ref{eq:diodPierce1}) with the
boundary conditions~(\ref{eq:diodPierce_usloviya}) describes the
dynamics of electron flow in the autonomous Pierce diode. It has been solved numerically
using finite difference approximation. Numerical solutions for
motion (\ref{eq:diodPierce1}) and continuity (\ref{eq:diodPierce2})
equations have been found by means of the explicit scheme with the
differences against flow, and Poisson equation
(\ref{eq:diodPierce3}) has been integrated using the error vector
propagation method \cite{Rouch:1976_FluidNumericalBook}. The time
and space integration steps have been selected as $\Delta x=0.005$
and $\Delta t=0.003$, respectively.

As a vector state of Pierce diode we have used the following one
\begin{equation}\label{eq:StateDefinition}
{\mathbf{U} = (\rho(x,t),v(x,t))^T}.
\end{equation}
In this case, according to~(\ref{eq:ScalarProductVectorfunction}),
the scalar product is defined as
\begin{equation}\label{eq:ScalarProduct}
(\mathbf{U_1},\mathbf{U_2}) = \int\limits_0^1
[\rho_1(x,t)\rho_2(x,t)+v_1(x,t)v_2(x,t)]\,dx,
\end{equation}
where the integration in~(\ref{eq:ScalarProduct}) should be performed
over a whole length of the considered system. Note, that the
potential $\varphi(x,t)$ is not included in the vector state
$\mathbf{U}$ since it could be explicitly defined through the
density of the electron beam $\rho(x,t)$ by Poisson
equation~(\ref{eq:diodPierce3}) and boundary
conditions~(\ref{eq:diodPierce_usloviya}). In other words, to
explicitly define the Pierce diode system state it is sufficient to
know the distributions of the density $\rho(x,t)$ of the electron
beam and velocity $v(x,t)$. It allows us, using the evolution operator
$\hat L(\mathbf{U})$, i.e. the
system~(\ref{eq:diodPierce3})-(\ref{eq:diodPierce1}) with boundary
conditions (\ref{eq:diodPierce_usloviya}), to calculate the
analogues states in the other moments of time.

As the perturbations $\mathbf{V}_i$ we have used the vectors
\begin{equation}\label{eq:Perturbation}
\mathbf{V}_i=(\xi^\rho_i(x,t),\xi^v_i(x,t))^T,
\end{equation}
characterizing the small deviations from the reference state
$\mathbf{U}$, with the small perturbation $\xi^\varphi_i(x,t)$ of the
potential $\varphi(x,t)$ being explicitly defined by the
distribution of the density of the electron beam $\rho(x,t)$ and its
small perturbation $\xi^\rho_i(x,t)$.

The behavior of Pierce diode with $i$-th perturbation of the state
$\mathbf{U}$ is described by the following system
\begin{equation}
\hat{L}(\mathbf{U}+\mathbf{V}_i)=0
\label{eq:PerturbatedEvolutionOperator}
\end{equation}
with the boundary conditions
\begin{equation}
\begin{array}{l}
 \rho(0,t)+\xi^\rho_i(0,t)=1, \\
 v(0,t)+\xi^v_i(0,t)=1,\\
 \varphi(0,t)+\xi^{\varphi}_i(0,t)=0,\\
 \varphi(1,t)+\xi^{\varphi}_i(1,t)=0.
\label{eq:PerturbedNetworkBoundary}
\end{array}
\end{equation}

Having linearized the system (\ref{eq:PerturbatedEvolutionOperator})
we can rewrite the equations
(\ref{eq:PerturbatedEvolutionOperator}),
(\ref{eq:PerturbedNetworkBoundary}) in the following way
\begin{equation}
\partial\hat{L}(\mathbf{U},\mathbf{V}_i)=0,
\label{eq:LinearEvolutionOperator}
\end{equation}
where $\partial\hat{L}(\mathbf{U},\mathbf{V}_i)$ is the evolution
operator $\hat{L}(\cdot)$ linearized near the reference state
$\mathbf{U}(x,t)$. It describes the character of the development of the
perturbations $\mathbf{V}_i$ with time. For the
system~(\ref{eq:diodPierce3})--(\ref{eq:diodPierce1}) the evolution
operator $\partial\hat{L}(\mathbf{U},\mathbf{V}_i)$ can be written
in the following way
\begin{equation}
\begin{array}{l}
\displaystyle\frac{\partial^2\xi_i^\varphi}{\partial x^2}=-\alpha^2\xi_i^\rho,\\
\\
\displaystyle\frac{\partial\xi_i^\rho}{\partial
t}=-\xi_i^\rho\frac{\partial v}{\partial x}-v\frac{\partial
\xi_i^\rho}{\partial x}-\xi_i^v\frac{\partial \rho}{\partial
x}-\rho\frac{\partial
\xi_i^v}{\partial x},\\
\\
\displaystyle\frac{\partial\xi_i^v}{\partial t}=-v\frac{\partial
\xi_i^v}{\partial x}-\xi_i^v\frac{\partial v}{\partial
x}-\frac{\partial \xi_i^\varphi}{\partial x}\\
\end{array}
\label{eq:LinearEqOfPerturbation}
\end{equation}
In this case the boundary
conditions~(\ref{eq:PerturbedNetworkBoundary}) for the small
deviations $\mathbf{V}_i$ of the synchronous state are given by
\begin{equation}
\begin{array}{l}
\xi^\rho_i(0,t)=0,\\
\xi^v_i(0,t)=0,\\
\xi^\varphi_i(0,t)=0,\\
\xi^\varphi_i(1,t)=0.
\end{array}
\label{eq:LinearBoundConditions}
\end{equation}
Obtained operator $\partial\hat{L}(\mathbf{U},\mathbf{V})$ is linear to
relatively small deviations $\mathbf{V}$.

Now, taking into account the reference state $\mathbf{U}$,
the perturbations $\mathbf{V}_i$, the evolution operator $\hat
L(\mathbf{U})$ defined for system~(\ref{eq:diodPierce3})--(\ref{eq:diodPierce1}) with boundary
conditions (\ref{eq:diodPierce_usloviya}), and the linearization
operator $\partial\hat{L}(\mathbf{U},\mathbf{V})$ given by
relations~(\ref{eq:LinearEqOfPerturbation})--(\ref{eq:LinearBoundConditions}),
we can calculate the spectrum of spatial Lyapunov exponents for the
autonomous model of Pierce diode.

The behavior of autonomous model of Pierce diode is known to be
defined by the only one dimensionless control parameter $\alpha$
called Pierce parameter. As we have mentioned above, Pierce diode is
capable to demonstrate the different types of the complex
spatio-temporal oscillations, with the transition from the periodic
to chaotic ones being performed via the period doubling bifurcation
cascade~\cite{Godfrey:1987, Kuhn:1990, Lindsay:1995, Matsumoto:1996,
Hramov:2004_IJE} when the $\alpha$-parameter value decreases in the
range $\alpha\in(2.862\pi,\,2.88\pi)$. Further decrease of the
parameter $\alpha$ results in the considerable complication of the
chaotic regime being realized in system
(\ref{eq:diodPierce3})-(\ref{eq:diodPierce1}) attended by the
disappearance of the well-defined time scale of the spatio-temporal
oscillations of the electron beam with the further transition to the
wide-band chaotic spatio-temporal oscillations.

\begin{figure}[tb]
\centerline{\scalebox{0.4}{\includegraphics{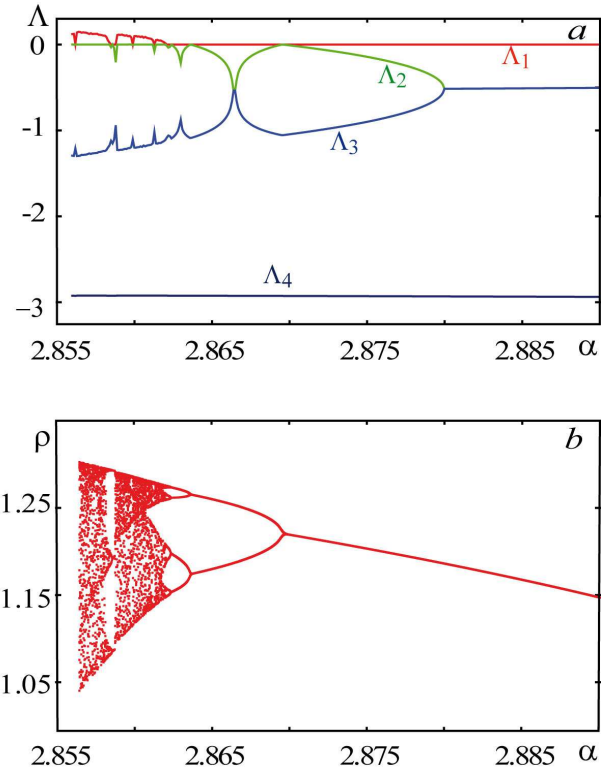}}}
\caption{(Color online) (\textit{a}) The dependence of the four highest Lyapunov exponents
and (\textit{b}) bifurcation diagram of oscillations in the
autonomous model of Pierce diode with $\alpha$-parameter value
increasing} \label{fgr:PierceDiode_LEsAndBif}
\end{figure}

To characterize the type of dynamical regime being realized in the
system under study we compute the spectrum of spatial Lyapunov
exponents. In Fig.~\ref{fgr:PierceDiode_LEsAndBif},\textit{a} the
dependence of the four highest spatial Lyapunov exponents on the
parameter $\alpha$ is shown. To calculate the values of the spectrum of the spatial Lyapunov exponents
$M=2\times10^4$ iterations of the renormalization and orthogonalizaiton have been done
for each value of the control parameter $\alpha$, with the time interval between iterations being equal to $T=0.6$.
In parallel, the bifurcation diagram of the density of the electron beam
oscillations at the point of the interaction space ${x=0.2}$ is also
indicated in Fig.~\ref{fgr:PierceDiode_LEsAndBif},\textit{b}.
It is easily seen from Fig.~\ref{fgr:PierceDiode_LEsAndBif} that the
spectrum of the highest spatial Lyapunov exponents allows to detect
clearly the type of the dynamical regime realized in the system under
study. For all values of the control parameter $\alpha$ one zero
Lyapunov exponent corresponding to the system perturbations along
the time axis is presented in the spectrum. In the region of the
periodic oscillations (e.g., ${\alpha\lesssim 2.862}$) such zero
Lyapunov exponent is the highest one, i.e., the chaotic dynamics is
absent. All other spatial Lyapunov exponents are negative. At the
same time, it is clearly seen that in the bifurcation points
corresponding to the period doubling cascade (see, e.g.,
$\alpha\simeq 2.87$) the second Lyapunov exponent $\Lambda_2$ is
also equal to zero, that indicates the qualitative changes in the
system behavior with the control parameter varying as well as for the
low-dimensional dynamical systems.

With the further decrease of the parameter $\alpha$ (see
Fig.~\ref{fgr:PierceDiode_LEsAndBif}) the highest spatial Lyapunov
exponent $\Lambda_1$ becomes positive, i.e. the transition to
chaotic oscillations takes place. The second Lyapunov exponent
remains to be the zero one. All others Lyapunov exponents are
negative. Note, that in the periodicity windows, where the periodic
oscillations take place, the positive Lyapunov exponent becomes also
negative and the zero Lyapunov exponent gets to be the
highest one.

It is important to note that in the case of the chaotic oscillations the
system under study has the only one positive Lyapunov exponent. At
the same time, the spatially extended media may theoretically have
several positive Lyapunov exponents. The presence of the only one
positive Lyapunov exponent is the evidence of the simplicity of
chaotic oscillations being realized in the spatially extended
system. One can assume that just due to that fact it has been
possible to construct the finite-dimensional model of the dynamics
of the electron beam in the Pierce diode being the system of
ordinary differential
equations~\cite{TrueAeh:2003_microwave_electronics_1engl} by means
of Galerkin method~\cite{Rouch:1976_FluidNumericalBook}.

So, the spatial Lyapunov exponents could be easily applied for the
detection and classification of the dynamical regimes being realized in the beam-plasma
systems with the overcritical current. However, the field of the
possible applications of the spectrum of the spatial Lyapunov
exponents is not restricted to the consideration of the autonomous
system. In particular, Lyapunov exponents are an effective tool for
the analysis of the different types of chaotic synchronization being
realized both in the finite- and infinite-dimensional
systems~\cite{Pyragas:1997_CLEsFromTimeSeries,Harmov:2005_GSOnset_EPL,Hramov:ZeroLE_PRE2008}.
Therefore, we consider the spectrum of spatial Lyapunov exponents
for two unidirectionally coupled Pierce diodes in the next
subsection.


\subsection{Two coupled Pierce diodes}

\label{sbsct:2PierceDiodes} The system under study in this case is given by the
partial differential equations
\begin{equation}
\frac{\partial^2\varphi^{1,2}}{\partial
x^2}=-\left(\alpha^{1,2}\right)^2(\rho^{1,2}-1),\label{Poisson_ch8}
\end{equation}
\begin{equation}
\frac{\partial \rho^{1,2}}{\partial t}=-\frac {\partial(\rho^{1,2}
v^{1,2})} {\partial x},\label{cotinuity_ch8}
\end{equation}
\begin{equation}
\frac{\partial v^{1,2}}{\partial t}= -v^{1,2} \frac{\partial
v^{1,2}}{\partial x}+\frac{\partial \varphi^{1,2}}{\partial
x},\label{motion_ch8}
\end{equation}
with the boundary conditions
\begin{equation}\label{q4_ch8}
v^{1,2}(0,t)=1,\quad \rho^{1,2}(0,t)=1,\quad\varphi^{1,2}(0,t)=0,
\end{equation}
where the top indexes ``1'' and ``2'' correspond to the first and second
coupled beam-plasma systems, respectively.

The unidirectional coupling between considered Pierce diodes is
realized by the modification of the boundary conditions on the right
boundary of the second (response) system, exactly in the same way as
it has been done in~\cite{Filatov:2006_PierceDiode_PLA}. In other words, the boundary
conditions for the system (\ref{Poisson_ch8})~--~(\ref{motion_ch8})
describing the dynamics of unidirectionally coupled Pierce diodes
should be written as
\begin{equation}
 \left \{
\begin{array}{lcl}
\varphi^1(1,t)&=&0\\
\varphi^2(1,t)&=&\varepsilon(\rho^2(x=1,t)-\rho^1(x=1,t)),\\
\end{array}
\right.\label{Coupl2}
\end{equation}
i.e., the first (drive ``1'') system is in the regime of autonomous
oscillations influencing on the second (response ``2'') one.

To analyze the behavior of two coupled Pierce diodes we compute spatial Lyapunov
exponents for the system (\ref{Poisson_ch8})~--~(\ref{motion_ch8}).
As a system state we use now the vector
\begin{equation}\label{eq:StateDefinition2}
{\mathbf{U} = (\rho^1(x,t),v^1(x,t),\rho^2(x,t),v^2(x,t))^T}.
\end{equation}
According to Eq.~(\ref{eq:ScalarProductVectorfunction}) the scalar product of the system states in this case is defined as
\begin{equation}\label{eq:ScalarProduct2}
\begin{array}{l}
(\mathbf{U_1},\mathbf{U_2}) = \int\limits_0^1
\left(\rho^1_1(x,t)\rho^1_2(x,t)+v^1_1(x,t)v^1_2(x,t)\right)dx+\\
\qquad \qquad \quad \int\limits_0^1
\left(\rho^2_1(x,t)\rho^2_2(x,t)+v^2_1(x,t)v^2_2(x,t)\right)dx.
\end{array}
\end{equation}
In Eq.~(\ref{eq:ScalarProduct2}) the integration is performed on the whole length of the
considered system as well as in the case of Eq.~(\ref{eq:ScalarProduct}).
Equations~(\ref{Poisson_ch8})--(\ref{motion_ch8}) with the boundary
conditions~(\ref{q4_ch8})--(\ref{Coupl2}) for two unidirectionally
coupled Pierce diodes play the role of the evolution
operator~(\ref{eq:Evolution}).

By analogy with the relation~(\ref{eq:Perturbation}) we use vectors
\begin{equation}\label{eq:Perturbation2}
\mathbf{V}_i=(\xi^{1\rho}_i(x,t),\xi^{1v}_i(x,t),\xi^{2\rho}_i(x,t),\xi^{2v}_i(x,t))^T,
\end{equation}
as the perturbations for two unidirectionally coupled Pierce diodes.
They characterize the small deviations from the reference state
$\mathbf{U}$, with the small perturbations $\xi^{1\varphi}_i(x,t)$
and $\xi^{2\varphi}_i(x,t)$ of the potentials $\varphi^{1,2}(x,t)$
as well as in the case of autonomous Pierce diode being explicitly
defined by the distributions of the density $\rho^{1,2}(x,t)$ of the
electron beam and its small perturbations  $\xi^{1,2\rho}_i(x,t)$.

For the system under study~(\ref{Poisson_ch8})--(\ref{Coupl2}) the
operator $\partial\hat{L}(\mathbf{U},\mathbf{V}_i)$ defining the
evolution of the small perturbations $\mathbf{V}_i$ of the reference
state $\mathbf{U}$ is written in the form
\begin{equation}
\begin{array}{ll}
\displaystyle\frac{\partial^2\xi_i^{1,2\varphi}}{\partial x^2} =& \displaystyle-\left(\alpha^{1,2}\right)^2\xi_i^{1,2\rho},\\
\\
\displaystyle\frac{\partial\xi_i^{1,2\rho}}{\partial t}
=&\displaystyle-\xi_i^{1,2\rho}\frac{\partial v^{1,2}}{\partial
x}-v^{1,2}\frac{\partial \xi_i^{1,2\rho}}{\partial
x}-\\&\displaystyle -\xi_i^{1,2v}\frac{\partial \rho^{1,2}}{\partial
x}-\rho^{1,2}\frac{\partial
\xi_i^{1,2v}}{\partial x},\\
\\
\displaystyle\frac{\partial\xi_i^{1,2v}}{\partial
t}=&\displaystyle-v^{1,2}\frac{\partial \xi_i^{1,2v}}{\partial
x}-\xi_i^{1,2v}\frac{\partial v^{1,2}}{\partial
x}-\frac{\partial \xi_i^{1,2\varphi}}{\partial x},\\
\end{array}
\label{eq:LinearEqOfPerturbation2}
\end{equation}
with the boundary conditions for the small deviations $\mathbf{V}_i$
from the reference state being the following
\begin{equation}
\begin{array}{l}
\xi^{1,2\rho}_i(0,t)=0,\\
\xi^{1,2v}_i(0,t)_i=0,\\
\xi^{1,2\varphi}_i(0,t)=0,\\
\xi^{1\varphi}_i(1,t)=0,\\
\xi^{2\varphi}_i(1,t)=\varepsilon(\xi^{2\rho}_i(1,t)-\xi^{1\rho}_i(1,t)).
\end{array}
\label{eq:LinearBoundConditionsForPerturbation}
\end{equation}

So, the computations of the spectrum of the spatial Lyapunov exponents
for two unidirectionally coupled Pierce diodes is performed in the
same way as for the autonomous system with the only one difference
consisting in the fact that the system state $\mathbf{U}$, small
perturbations $\mathbf{V}_i$, the scalar product
$(\mathbf{U}_1,\mathbf{U}_2)$ and the evolution operators $\hat
L(\mathbf{U})$ and $\partial\hat L(\mathbf{U},\mathbf{V}_i)$ are
defined for two interacting systems.

\begin{figure}[tb]
\centerline{\scalebox{0.4}{\includegraphics{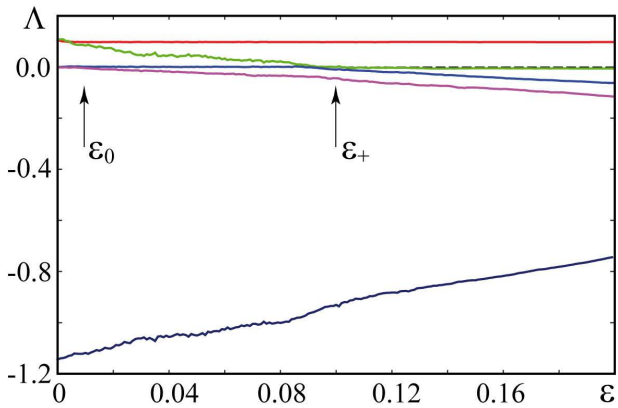}}}
\caption{(Color online) The dependence of the five highest Lyapunov exponents on the
coupling parameter $\varepsilon$ for two unidirectionally coupled
Pierce diodes} \label{fgr:UniCoupledPierceDiodes_LEs}
\end{figure}

The dependence of the five highest spatial Lyapunov exponents on the
coupling parameter strength $\varepsilon$ for two unidirectionally
coupled Pierce diodes~(\ref{Poisson_ch8})-(\ref{motion_ch8}) with
the boundary conditions~(\ref{q4_ch8})--(\ref{Coupl2}) is shown in
Fig.~\ref{fgr:UniCoupledPierceDiodes_LEs}. The control parameter
values $\alpha^1=2.858$, $\alpha^2=2.860$ have been chosen in such a
way that the autonomous dynamics of each Pierce diode is being
chaotic one. It is clearly seen that in the absence of the coupling
$\varepsilon=0$ two highest Lyapunov exponents are positive (that is
the evidence of the presence of the chaotic dynamics in each Pierce
diode), two Lyapunov exponents are equal to zero (that corresponds
to the perturbations of the system states along the time axis), the
fifth and others Lyapunov exponents are negative.

With the coupling parameter value increase the quantitative values
of the Lyapunov exponents start changing. Due to the unidirectional
type of the coupling between Pierce diodes the spectrum of Lyapunov
exponents ${\Lambda_1\geq\Lambda_2\geq\dots}$ can be divided into
two parts. One of them corresponds to the Lyapunov exponents
calculated for the drive system
${\Lambda^1_1\geq\Lambda^1_2\geq\dots}$ and due to the independency of
the behavior of the drive system on the response one they do not
depend on the coupling parameter strength. The second part is the
conditional Lyapunov exponents, i.e., Lyapunov exponents computed
for the response system. They depend on the coupling parameter
strength $\varepsilon$ and, therefore, are called conditional.

Such a model of Pierce diode is known to demonstrate the different
regimes of chaotic synchronization, including complete, generalized
and time scale synchronization~\cite{Filatov:2006_PierceDiode_PLA}. The
occurrence of any type of the synchronous regime should be
accompanied by the transformations in the spectrum of Lyapunov
exponents. In particular, for the systems with a small number of
degrees of freedom the transition of the zero conditional
(corresponding to the response system) Lyapunov exponent in the
field of the negative values precedes the phase or time scale
synchronization regimes in the case of the small values of the
parameter detuning~\cite{Rosenblum:1997_LagSynchro,Hramov:2007_2TypesPSDestruction,Hramov:ZeroLE_PRE2008}. When the highest conditional Lyapunov
exponent passes through zero, the generalized synchronization regime
takes place~\cite{Pyragas:1996_WeakAndStrongSynchro,Filatov:2006_PierceDiode_PLA}.

The very same phenomena is observed for two coupled spatially extended beam plasma systems.
When the coupling parameter value increases the response system becomes
synchronized by the drive one, with the qualitative changes in the
spectrum of the conditional Lyapunov exponents taking place. In
Fig.~\ref{fgr:UniCoupledPierceDiodes_LEs} two critical points
$\varepsilon_0\approx0.01$ and $\varepsilon_{+}\approx0.09$ are clearly seen.
As in the case of the low-dimensional system, the first of them
corresponds to the transition of the zero conditional Lyapunov exponent
in the field of the negative values, whereas the second one relates
to the zero-pass of the positive conditional Lyapunov exponent.
The negativeness of the largest conditional spatial Lyapunov exponent is the evidence of the generalized synchronization regime that has been confirmed with the help of the auxiliary system method~\cite{Rulkov:1996_AuxiliarySystem,Filatov:2006_PierceDiode_PLA}.


\section{Spectrum of spatial Lyapunov exponents for the transverse field backward wave oscillator}
\label{Sct:Backward wave oscillator}
To illustrate the generality and efficiency of the proposed method, as well as their applicability for the
widespread class of  spatially-extended systems, we consider the behavior of  the transverse field backward wave oscillator
being the classical models of the electrons-wave systems, demonstrating the complex oscillations including the chaotic regimes as well as hyperchaos.

The dynamics of the backward wave oscillator is described by the system of the dimensionless equations
\begin{equation}\label{eq:LOV1}
\frac{\partial F}{\partial t}- \frac{\partial F}{\partial
x}=-AI,
\end{equation}
\begin{equation}\label{eq:LOV2}
\frac{\partial I}{\partial x}+ j\mid I \mid^2 I =-AF,
\end{equation}
where $F(x,t)$ is the dimensionless complex-valued amplitude of the electric field and $I(x,t)$ is the complex-valued current $(0<x<1)$. The control parameter $A$ can be considered as  the dimensionless length of the system. The equations~(\ref{eq:LOV1})-(\ref{eq:LOV2}) have been integrated
with the boundary  conditions:
\begin{equation}
F(1,t)=0,\quad I(0,t)=0.
\label{eq:BoundaryLOV}
\end{equation}

To calculate the spectrum of  Lyapunov exponents we define the main state of this system as
\begin{equation}\label{eq:StateLOV}
{\mathbf{U} = F(x,t)}.
\end{equation}
It is important to note, that the distribution of the complex-valued current is excluded from the main system state, since it is explicitly defined by the electrical field $F(x,t)$.

The scalar product of complex-valued functions in this case is defined as follows
\begin{equation}\label{eq:ScalarProductLOV}
(\mathbf{U_1},\mathbf{U_2}) = \int\limits_0^1
F_1(x,t)F_2^*(x,t)\,\,dx,
\end{equation}
where the integration, as for the case of  Pierce diode, should be performed over the whole length of the considered system.

The main system state~(\ref{eq:StateLOV}) evaluates according to the operator ~(\ref{eq:Evolution}), presented in the form of differential equations~(\ref{eq:LOV1})-(\ref{eq:LOV2}). These equations~(\ref{eq:LOV1})-(\ref{eq:LOV2}) have been integrated using the Lax-Wendroff method and the third-order Runge-Kutta method. The time and space integration steps have been selected as  $\Delta x= 0.004 $ and $\Delta t= 0.002$, respectively.

The state~(\ref{eq:StateLOV}) plays the role of the reference state. To calculate $N$ largest spatial Lyapunov exponents we consider the set of perturbations
\begin{equation}\label{eq:PerturbationLOV}
\mathbf{V}_i=(\xi^F_i(x,t)),
\end{equation}
Again, the current has been eliminated from the consideration,
since the small perturbations  $\xi_i^I(x,t)$ of the complex-valued current $I(x,t)$ are explicitly defined by the distribution of the high-frequency electric field $F(x,t)$ and its small perturbations $\xi_i^F(x,t)$.

The evolution of the set of small perturbations is described by the linearized operator~(\ref{eq:PerturbationEvolution}), which for the considered system can be written in the form:
\begin{equation}
\begin{array}{l}
\displaystyle\frac{\partial \xi^F_i}{\partial t}- \frac{\partial \xi^F_i}{\partial x}=-A\xi^I_i,\\
\displaystyle\frac{\partial\xi^I_i}{\partial t}+2j\mid I \mid^2 {\xi^I_i}^* + j I ^2 \xi^I_i=-A\xi^F_i\\
\end{array}
\label{eq:LinearEqOfPerturbationLOV}
\end{equation}
with the boundary conditions
\begin{equation}
F(1,t)+\xi^F_i(1,t)=0,\quad I(0,t)+\xi^I_i(0,t)=0.
\label{eq:PerturbedNetworkBoundaryLOV}
\end{equation}
The Lyapunov sums and values of SLEs have been estimated with the help of expressions~(\ref{eq:SLEsums})-(\ref{eq:SLE}).

\begin{figure}[tb]
\centerline{\scalebox{0.4}{\includegraphics{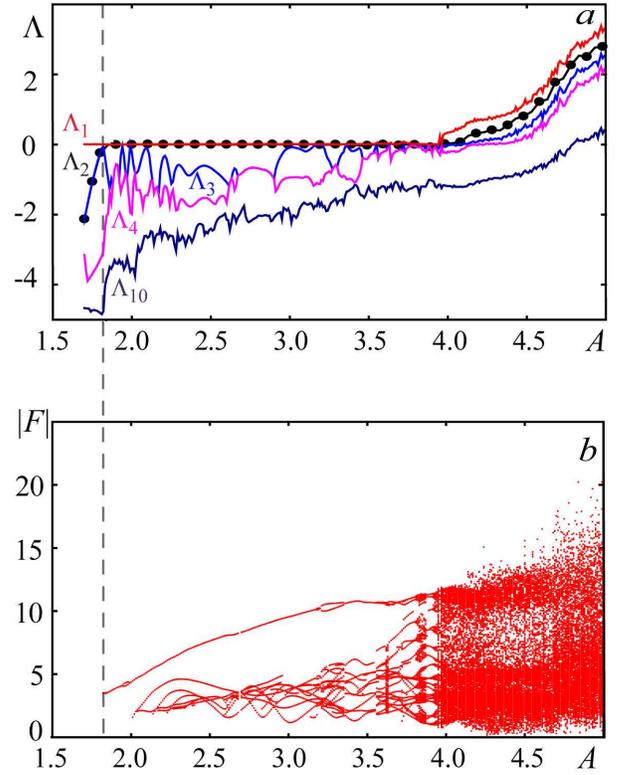}}}
\caption{(Color online) (\textit{a}) The dependence of the four highest and the 10th Lyapunov exponents (the second LE is shown by points ($\bullet$))
and (\textit{b}) the bifurcation diagram of oscillations in the transverse field Backward wave oscillator with the $A$-parameter value
increase} \label{fgr:LOV_bif_LEs}
\end{figure}

In Fig.~\ref{fgr:LOV_bif_LEs},\textit{a} the
dependence of the four highest spatial Lyapunov exponents on the
parameter $A$ is shown. To give the information of the behavior of the lower SLEs the dependence of the 10th LE is also given in Fig.~\ref{fgr:LOV_bif_LEs},\textit{a}.
The bifurcation diagram of the complex-valued amplitude of the electric field
is presented in Fig.~\ref{fgr:LOV_bif_LEs},\textit{b}.
The largest Lyapunov exponent $\lambda_1$ is equal to zero when the control parameter has the value being less then $1.83$, that corresponds to the stationary state, when the antiphase oscillations of the real and imaginary parts of the complex-valued amplitude of electric field takes place, while its module remains constant. All other SLEs $\lambda_2,\lambda_3,\dots$ are negative.

For the bifurcation point $A_c\approx 1.83$ the stationary state becomes unstable and, as a consequence, the periodic oscillations of the module of the complex-valued electric field $|F|$ are observed in the system under study. In the region of the periodic oscillations $A>A_c$ the second Lyapunov exponent $\lambda_2$ is also equal to zero. At the same time, all other Lyapunov exponents are negative, as before. Note also, at the bifurcation point $A_c$ the third Lyapunov exponent $\lambda_3$ is equal to zero, that indicates the qualitative changes taking place in the spatially-extended system.

With the further increase of the $A$-parameter the transition from periodic oscillations to chaotic dynamics is observed. This transformation is accompanied by the modification of the spectrum of SLEs, when the largest Lyapunov exponent becomes positive ($A_{ch}\approx 3.9$). As for the case of Pierce diode, it can be used as a criterium of the transition to the chaotic regime, and, as well as for Pierce diode, this transformation of the SLEs spectrum agrees with the bifurcation diagram (Fig.~\ref{fgr:LOV_bif_LEs},\textit{a}).

As it has been described above, the chaotic regimes in the spatially-extendent systems, in general, can be characterized by several positive Lyapunov exponents. It is clearly seen from Fig.~\ref{fgr:LOV_bif_LEs},\textit{b}, the number of the positive Lyapunov exponents increases with the grow of the value of $A$-parameter, with the chaotic regime (see bifurcation diagram, Fig.~\ref{fgr:LOV_bif_LEs},\textit{a}) becoming more complicated. Note also, as soon as one of the Lyapunov exponents being zero becomes positive, the largest negative Lyapunov exponent takes the value zero.

\section{The selection of the system state: brief discussion}
\label{sct:SystemStateSelection}

The proposed approach for the computation of the spectrum of the spatial Lyapunov exponents is shown to be the effective and powerful tool to study and analyze the complex behavior of the beam plasma systems as well as the other spatially extended media. Therefore, it is not surprised that this tool attracts great attention of scientists~\cite{Kuznetsov:2004_LEinBWO,Hramov:2008_INIS_PRE}. Nevertheless, one have to be very careful computing the spectrum of the spatial Lyapunov exponents, since there may be certain particularities connected with the SLEs calculation which may result to the inaccurate or even incorrect decisions made and, therefore, they must be taken into account by scientists. In this section one of the possible problems will be discussed for the scientists to avoid the potential errors.

When the spectrum of spatial Lyapunov exponents is calculated \emph{the state} $\mathbf{U}$ of the system should be chosen properly. In the studies given above certain variables (such as the potential $\varphi(x,t)$ in the case of Pierce diode or the current $I(x,t)$ in the case of the transverse field backward wave oscillator) have been excluded from the system state having based on the reason that these variables may be explicitly defined through the other ones which are included in the system state. To prove the validity of this approach we have also calculated the spectrum of the spatial Lyapunov exponents when the whole set of the variables is supposed to be the system state. For the Pierce diode the system state in this case is
\begin{equation}\label{eq:FullStateDefinition}
{\mathbf{U} = (\varphi(x,t),\rho(x,t),v(x,t))^T},
\end{equation}
whereas for the transverse field backward wave oscillator it is defined as
\begin{equation}\label{eq:FullStateLOV}
{\mathbf{U} = (F(x,t),I(x,t))^T}.
\end{equation}

\begin{figure}[tb]
\centerline{\scalebox{0.35}{\includegraphics{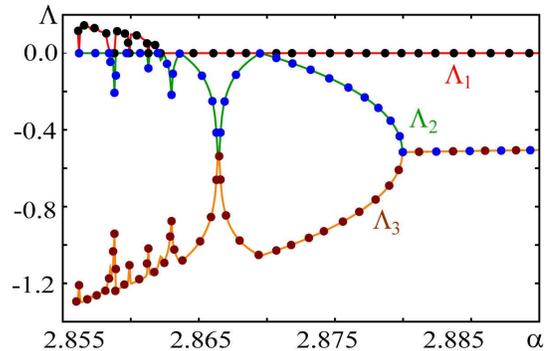}}}
\caption{(Color online) The three highest Lyapunov exponents of Pierce diode
calculated both for $\mathbf{U}=(\varphi(x,t),\rho(x,t),v(x,t))^T$ (points) and for $\mathbf{U}=(\rho(x,t),v(x,t))^T$ (lines)
} \label{fgr:Pirs_LES_2}
\end{figure}

\begin{figure}[tb]
\centerline{\scalebox{0.4}{\includegraphics{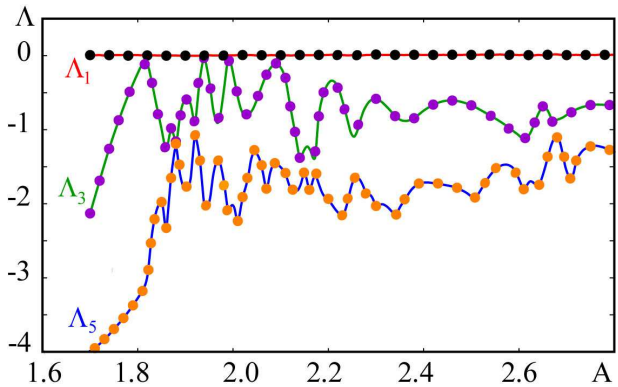}}}
\caption{(Color online) The first, the third and the fifth highest Lyapunov exponents of the backward wave oscillator
calculated both for $\mathbf{U} = (F(x,t),I(x,t))^T$ (points) and $\mathbf{U}=F(x,t)$ (lines)} \label{fgr:LOV_LES_2}
\end{figure}

Fig.~\ref{fgr:Pirs_LES_2} illustrates the three highest Lyapunov exponents of Pierce diode obtained in both cases (when $\mathbf{U}=(\varphi(x,t),\rho(x,t),v(x,t))^T$ and $\mathbf{U}=(\rho(x,t),v(x,t))^T$). One can see that the LEs correspond exactly to the values, calculated in Sec.~\ref{sct:LEsForPierceDiode}. Analogously, in Fig.~\ref{fgr:LOV_LES_2} the first, the third and the fifth Lyapunov exponents corresponding to the periodical oscillations in the backward wave oscillator are shown. It is clearly seen, that the values of Lyapunov exponents obtained in the both cases ($\mathbf{U}=(F(x,t),I(x,t))^T$ and $\mathbf{U}=F(x,t)$) are identical for all values of the dimensionless system length $A$.

So, certain variables can be excluded from the system state if they may be explicitly defined through the other system variables in the every moment of time or, in other words, if the time derivatives of these variables are not presented in the evolution operator. Nevertheless, in general, the choice of the system state $\mathbf{U}$  requires the thoroughness and accuracy. Thus, the majority of the beam plasma and microwave systems are simulated with the help of the particle methods~\cite{Hockney:1981}. In this case the state $\mathbf{U}(x,t)$ of the system under study is determined by the electromagnetic field as well as locations and velocities of all particles used for simulation. As far as the electromagnetic field is concerned there is no problem to apply the renormalization and  orthogonalizaiton procedures in the very same way as it has been described in Sec.~\ref{sct:SLEsTheory}, since this field is the function of the coordinate, i.e., $E(x,t)\in\mathbb{L}^2$, $B(x,t)\in\mathbb{L}^2$. On the contrary, the set of particles emulating the electron beam can not be renormalized and orthogonalized with the help of the technique given above, since each particle is characterized by its own coordinate and velocity. As a consequence, the scientists dealing with such a type of systems use only the electromagnetic field as the
system state $\mathbf{U}(x,t)$ to avoid the uncertainty connected with particle dynamics evaluation (see, e.g.~\cite{Kuznetsov:2004_LEinBWO}).
Although in some instances the obtained in such a way results seem to be plausible, we show below that this approach gives the very crude and inaccurate solution.

To illustrate this point we consider the emulation of the spatial Lyapunov exponent calculation for the spatially extended system simulated by the particle method by the example of Pierce diode. Since in this case the electron beam (``particles'') can not be used when the renormalization and  orthogonalizaiton procedures are implemented, we consider the truncated system state
\begin{equation}\label{eq:PierceDiodeTruncState}
\mathbf{U}'(x,t)=v(x,t)
\end{equation}
in parallel with~(\ref{eq:StateDefinition}). The dynamics of Pierce diode state and its perturbations are evaluated according to the operator~(\ref{eq:diodPierce3})-(\ref{eq:diodPierce1}), whereas the Gramm-Schmidt procedure is realized only for the truncated state~(\ref{eq:PierceDiodeTruncState}) that excludes the beam of electrons from the renormalization and  orthogonalizaiton, exactly in the same way, as it has been done, e.g., in~\cite{Kuznetsov:2004_LEinBWO} for the system simulated by the particle method.

\begin{figure}[tb]
\centerline{\scalebox{0.4}{\includegraphics{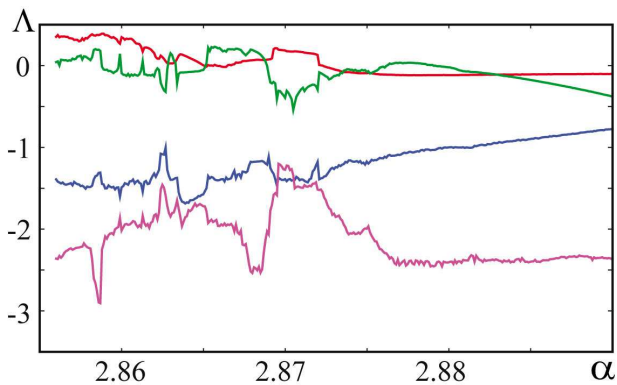}}}
\caption{(Color online) The four highest Lyapunov exponents
calculated for the case when the beam of electrons is excluded from the renormalization and  orthogonalizaiton procedure (compare with Fig.~\ref{fgr:PierceDiode_LEsAndBif},\,\textit{a})} \label{fgr:PierceDiode_LEsTruncated}
\end{figure}

The four largest Lyapunov exponents (solid lines) calculated for Pierce diode by means of the described technique (it models the approach used for the system simulated by particle method) are shown in Fig.~\ref{fgr:PierceDiode_LEsTruncated} to be compared with the Lyapunov exponents calculated in Sec.~\ref{Sct:PierceDiode} (see Fig.~\ref{fgr:PierceDiode_LEsAndBif}). It is easy to see that the true dependence of the spatial Lyapunov exponents (Fig.~\ref{fgr:PierceDiode_LEsAndBif}) is seriously distorted (Fig.~\ref{fgr:PierceDiode_LEsTruncated}) when the electron beam is excluded from the renormalization and  orthogonalizaiton procedures. E.g., even zero Lyapunov exponent being known to have always to exist in the autonomous system (since it corresponds to the perturbation along with the time axis) is not calculated properly in comparison with the case, when all sufficient quantities (in the considered case these are $\rho(x,t)$ and $v(x,t)$) are taken into account.

Evidently, for some values of the control parameter $\alpha$ the obtained set of spatial Lyapunov exponents seems to be reasonable. E.g., for $\alpha=2.856$ the largest value of the calculated spatial Lyapunov exponent spectrum is positive that corresponds to the chaotic dynamics observed in the system, the second Lyapunov exponent is close to zero that also agrees well with the theory, the other Lyapunov exponents are negative. So, if the careful verification of the obtained results has not been done these results may be assumed to be correct, although the values of the calculated spatial Lyapunov exponents differ greatly from the true ones. Moreover, for the other values of the control parameters the calculated data may be obviously contradictory (see, e.g., ${\alpha=2.87}$, where despite of the periodic regime the positive value of the spatial Lyapunov exponent is obtained, whereas the zero Lyapunov exponent is absent). The inconsistency of the obtained values of the spatial Lyapunov exponents with the system dynamics may be revealed with the help of the examination of the dependence of SLEs on the system control parameter value, but, typically, this analysis is not given.

Therefore, we can come to the conclusion that the calculation of the spatial Lyapunov exponents for the models simulated by the particle methods must be used very carefully. Unfortunately, at present there is no methods allowing the calculations of SLEs for such kind of the dynamical systems and this problem should be solved in the future.

\section*{Conclusion}

In the present work we have proposed the method for the calculation of the spectrum of the spatial Lyapunov exponents for the spatially extended beam plasma systems. This method is shown to be effective and powerful tool for the analysis of the complex behavior of the beam plasma systems as well as the other spatially extended systems. The possible applications of the spectrum of the spatial Lyapunov exponents are not restricted to the consideration of the autonomous system, e.g., SLEs are the effective tool for
the analysis of the different types of chaotic synchronization. At the same time, the calculation of the spatial Lyapunov exponents for the models simulated by the particle methods must be used very carefully, whereas the problem of calculations of SLEs for these systems should be solved in the future.

Though the calculation of the spectrum of the spatial Lyapunov exponents has been illustrated with the help of several examples, we expect that this technique may be used in many other relevant circumstances,  as
e.g. laser systems~\cite{Boccaletti:2002_LaserPSTransition_PRL}, semiconductor superlattices~\cite{Balanov:2009_SL_PRB}, etc.


\end{document}